# Converse flexoelectricity around ferroelastic domain walls

Y. J. Wang[1]†, Y. L. Tang[1]†, Y. L. Zhu[1]*, Y. P. Feng[1,2], and X. L. Ma[1,3]*

[1]*Shenyang National Laboratory for Materials Science, Institute of Metal Research, Chinese Academy of Sciences, Wenhua Road 72, 110016 Shenyang, China*
[2]*University of Chinese Academy of Sciences, Yuquan Road 19, 100049 Beijing, China*
[3]*State Key Lab of Advanced Processing and Recycling on Non-ferrous Metals, Lanzhou University of Technology, Langongping Road 287, 730050, Lanzhou, China*

†Authors contributed equally in this work.
*Correspondence should be addressed to X. L. Ma (xlma@imr.ac.cn) or Y. L. Zhu (ylzhu@imr.ac.cn)

Domain walls (DWs) are ubiquitous in ferroelectric materials. Ferroelastic DWs refer to those who separate two domains with unparalleled polarizations (or two different ferroelastic variants). These ultrathin interfaces could be designed as minimized electronic devices[1] since they possess fascinating functionalities, such as conductivity[2], abnormal photovoltaic effect[3], and so on. They are responsible to the enhancement of dielectric and piezoelectric properties of ferroelectric materials[4,5]. It is long believed that the structures of ferroelastic DWs can be simply explained from the perspective of mechanical and electric compatibilities in the framework of the Landau-Ginzburg-Devonshire (LGD) theory. Here we show that the converse flexoelectricity must be taken into account for fully describing the nature of ferroelastic DWs. The converse flexoelectricity is a stress or a strain induced by polarization gradients[6], which is complementary to the relatively well-known direct flexoelectricity (the polarization induced by strain gradients)[7]. In our work, an unexpected asymmetric structure is identified, which is beyond the prediction of the conventional LGD theory. By incorporating the converse flexoelectricity into the LGD theory and using it to analyze high-resolution images acquired by the aberration-corrected transmission electron microscope (TEM), we demonstrate that it is the converse flexoelectricity that result in the asymmetric structure. Moreover, the flexoelectric coefficient is derived by quantifying the converse flexoelectricity around the DWs. This quantification is deterministic in both the magnitude and sign of flexoelectric coefficients, by the mutual verification of atomic mapping and first-principles calculations. Our results suggest that the converse flexoelectricity cannot be neglected for understanding the ferroelastic DWs and other boundaries in ferroelectric materials.

DWs are the physical separation of two domains with different polarization vectors in ferroelectric materials. DWs attract great attention due to their abundant functionalities[2,3], which are closely related to their internal structures. For example, DW conductivity was thought to be related with the normal polarization components induced by the direct flexoelectric effect[8]. DWs can be classified into two categories according to the polarization vectors of the two separated domains: if they are antiparallel to each other, a ferroelectric 180°DW is formed; if they are unparalleled, however, a ferroelastic DW is formed. Ferroelastic DWs can mediate misfit strain or external stress, thus, playing important roles in the formation of domain structures[9]. Besides, they are also found to enhance the dielectric and piezoelectric properties of ferroelectric materials[4,5]. The structures of ferroelastic DWs were previously thought to be simply determined by both the mechanical and electric compatibilities. As a result, ferroelastic DWs usually lie on specific crystallographic planes, a feature different from ferroelectric 180°DWs. Recently, an asymmetric structure was found around 90°DWs in tetragonal $PbTiO_3$ (PTO)[10], which cannot be explained by the previous wisdom.

To explain this asymmetric structure, TEM experiments were carried out to reproduce this result and obtain detailed atomic-scale information. A PTO film of 200 nm was grown on a $(001)_{PC}$-oriented $DyScO_3$ ($DSO(001)_{PC}$, 'PC' denotes the pseudocubic lattice) substrate. Stripe *a/c* domains were observed in the cross-sectional image (Supplementary Fig. 1). From the high-angle annular dark-field (HAADF) image in Fig. 1a, two 90°DWs can be clearly identified from the large lattice rotation, as marked by yellow dashed lines. Peaks and valleys can be clearly revealed in the averaged lattice parameter profiles as denoted by vertical arrows (Fig. 1b). These results indicate that the transition of lattice parameters across 90 °DWs is not monotonous but discontinuous. The observed asymmetric structure around 90 °DWs is nearly the same as that in the previous study[10].

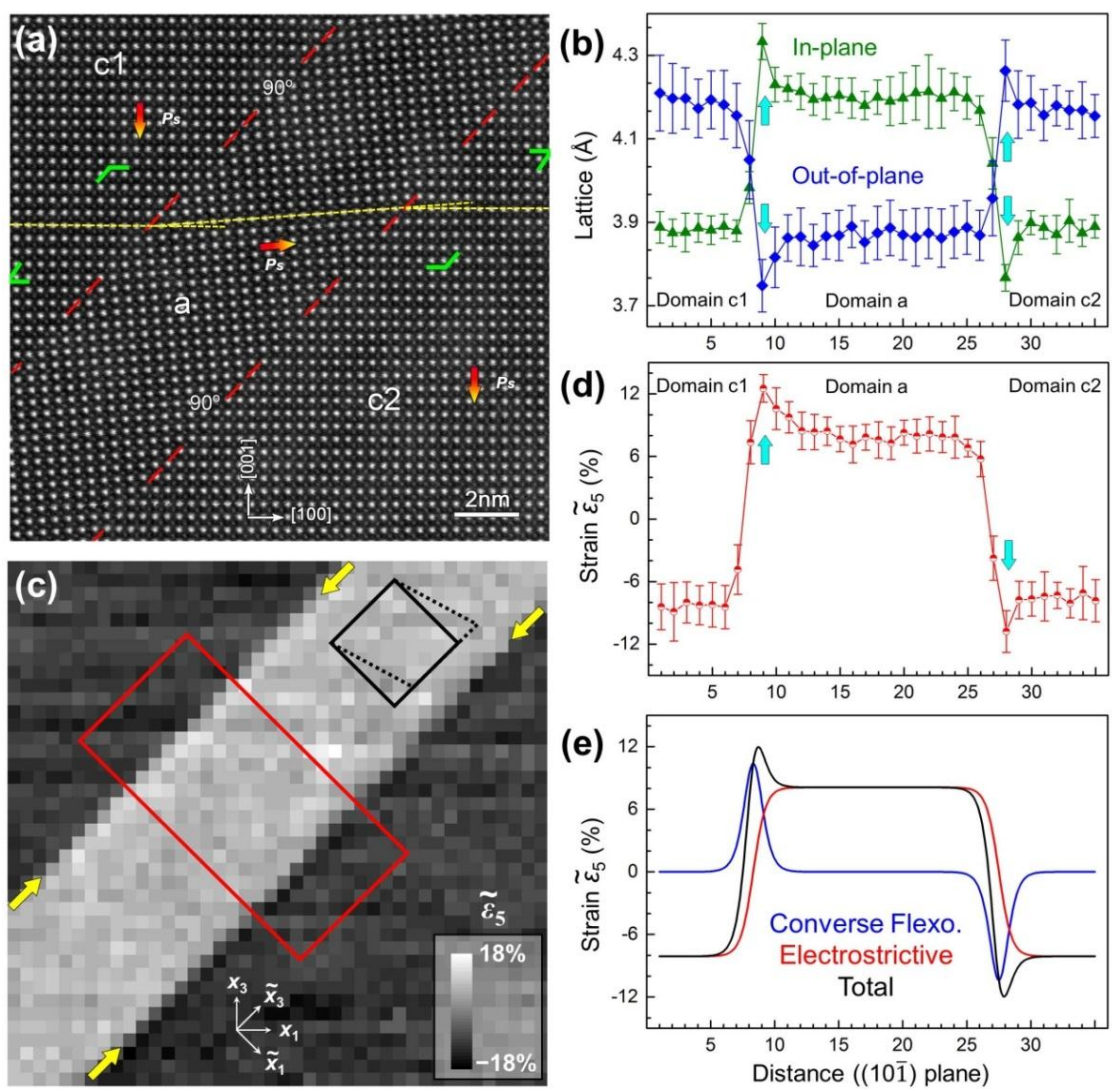

**Figure 1 | The asymmetric structure around 90 °DWs in PTO films due to the converse flexoelectricity.** (**a**) A high-resolution HAADF image of 90 °DWs in a PTO film grown on DSO(001)$_{PC}$. Three domains were labeled as c1, a and c2. Two 90 ° domain walls were marked with red dotted lines. (**b**) Averaged lattice parameters across the 90 °DWs. The area labeled with green markers in **a** was used as the data source. (**c**) Mappings of shear strain $\tilde{\varepsilon}_5$. (**d**) Averaged shear strain across the 90 °DWs. The area labeled with red boxes in **c** was used as the data source. (**e**) The fitting curve of **d** and the decomposition of the shear strain. The fitting function and parameters are given in Supplementary Note 2.

Notice that the variations of lattice parameters along [100] and [001] directions correspond to normal strains in $x_1$ and $x_3$ directions ($\varepsilon_1$ and $\varepsilon_3$). By doing the coordinate transformation (see Supplementary Fig. 2 and Supplementary Note 1), the shear strain in the $(10\bar{1})$ plane (the DW plane) $\tilde{\varepsilon}_5 = 2\tilde{\varepsilon}_{13} = \varepsilon_1 - \varepsilon_3$ was calculated (Fig. 1c). Symbols with and without tildes correspond to the physical quantities in different coordinate systems, as illustrated in Fig. 1c. A peak and a valley were found in the averaged profile of $\tilde{\varepsilon}_5$ as denoted by vertical arrows (Fig. 1d). Comparing Figs. 1b and 1d, it is found that the asymmetric structure in the lattice parameter profile is equivalent to the one in the shear strain profile.

According to the LGD theory incorporating the converse flexoelectricity (Supplementary Note 1[8,11-14]), the shear strain $\tilde{\varepsilon}_5$ around a 90 °DW is:

$$\tilde{\varepsilon}_5 = -\tilde{F}_{55}\frac{d\tilde{P}_3}{d\tilde{x}_1} + \tilde{Q}_{55}\tilde{P}_1\tilde{P}_3 \tag{1}$$

where $\tilde{P}_1$ and $\tilde{P}_3$ are the polarization components normal and parallel to the DW, respectively; $\tilde{F}_{55} = F_{11} - F_{12}$ and $\tilde{Q}_{55} = Q_{11} - Q_{12}$ are flexoelectric and electrostrictive coefficients. From Eq. (1), the shear strain can be decomposed into two parts: the one related to the converse flexoelectric effect and the other one related to the electrostrictive effect (Fig. 1e and Supplementary Note 2). If only the electrostrictive effect is considered, the shear strain profile should be smooth as shown with the red line in Fig. 1e. It is the converse flexoelectric effect that causes the peak and valley in the shear strain profile, that is, the asymmetric structure at 90 °DWs.

From Eq. (1), it is indicated that the flexoelectric coefficient $F_{11}-F_{12}$ can be determined by fitting Eq. (1). Actually, the fitting is more convenient by integrating Eq. (1) to obtain the displacement function (Supplementary Note 1). To obtain an accurate coefficient requires that the 90 °DW should be very straight. The 90 °DWs in the (001)-oriented PTO film does not meet this requirement (Steps were observed along 90 ° DWs). Thus, a PTO film of 50 nm was grown on a (101)-oriented (La,Sr)(Al,Ta)O$_3$ (LSAT(101)) substrate and 90 °DWs perpendicular to the film/substrate interface were found (Supplementary Fig. 3), similar to our recent finding[15]. In this orientation, the two domains beside a 90 °DW are in symmetric strain states, so that most 90 °DWs are very straight. Similar to Fig. 1a, a lattice rotation across the 90 °DW can be clearly seen, as marked by the red dashed line (Fig. 2a). The displacements of Pb ions were extracted from the HAADF image (Fig. 2b) and from the fitting result of the zoom-in image (Fig. 2c), it is found that there is a Pb sublattice offset across the DW (about 0.71 Å). If Fig. 2a is rotated 45 ° clockwise, the polarization configuration around this 90 ° DW is exactly the same as the bottom right one in Fig. 1a. As was expected, a valley is found in the shear strain profile (Fig. 2d).

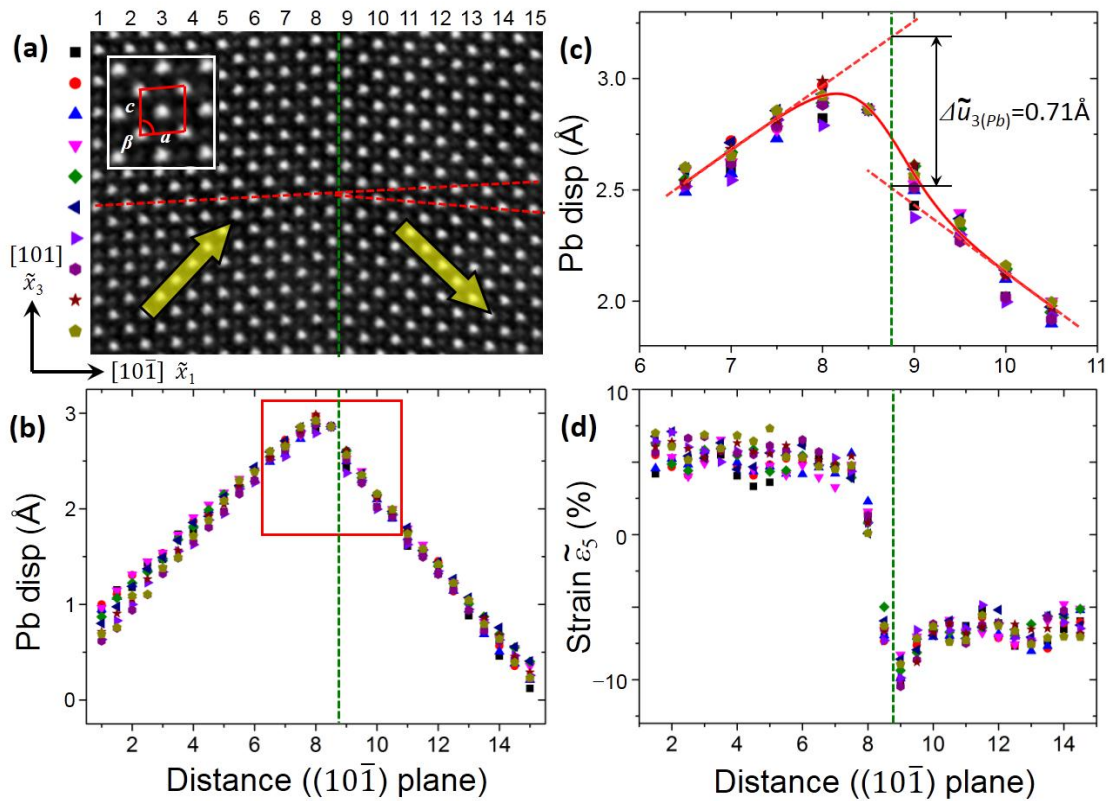

**Figure 2 | Pb sublattice offset across a 90 °DW in a (101)-oriented PTO film.** (**a**) The high-resolution HAADF image of a 90 °DW in a PTO film grown on LSAT(101). (**b**) The Pb atomic displacements in the $\tilde{x}_3$ direction. (**c**) The zoom-in image of the red rectangle in **b**. The solid line in **c** is the fitting curve. The fitting function is given in Supplementary Note 1. The Pb sublattice offset is labeled in **c**. (**d**) The shear strain $\tilde{\varepsilon}_5$ = cot($\beta$) around the 90 °DW. The green dashed lines mark the DW plane according to the fitting result.

The displacements shown in Fig. 2 contain only the contribution of the Pb sublattice. The contributions of Ti and O sublattices should also be considered to obtain the authentic displacements. Thus, first-principles calculations were performed to determine the coordinates of all atoms. From the HAADF image, the lattice parameters of a unit cell away from the DW (the insert in Fig. 2a) is determined to be $a = 5.703$ Å, $c = 5.720$ Å, and $\beta = 86°$. The lattice parameter in the direction of the electron beam is assumed to be the same as the substrate, $b = 3.869$ Å. Based on these experimental lattice parameters (ELP), the atomic model of 90° DWs was constructed (Fig. 3a). Besides this model, another DW model was built from calculated lattice parameters (CLP). Two types of exchange-correlation (EC) functionals (local density approximation (LDA) and generalized gradient approximation for solids (PBE-sol)) were adopted in the calculations and other technical details can be found in the method section, Supplementary Fig. 4, and Supplementary Note 3[16-18]. The Pb sublattice shows an offset of about 0.61 Å across the DW (Fig. 3b), which is consistent with the experimental value. Besides this consistence, peaks and valleys can also be found in the calculated shear strain profile (Fig. 3c). Considering the contribution of all atoms, the displacements of unit cells are calculated (Fig. 3b) by the method illustrated in Supplementary Figs. 5 and 6 and Supplementary Note 4. The offsets of the Pb sublattice and unit cells are calculated by fitting the displacements, similar to Fig. 2c. It is found that the offset of unit cells is smaller than that of the Pb sublattice. The unit cell's offset $\Delta \tilde{u}_3$ and the spontaneous polarization $P_s$, which is calculated by the Berry phase method[19], are linked by the following formula (Supplementary Note 1):

$$\Delta \tilde{u}_3 = -\sqrt{2} \tilde{F}_{55} P_s \quad (2)$$

From this formula, the flexoelectric coefficient $\tilde{F}_{55} = F_{11} - F_{12}$ can be calculated as $-3.5 \times 10^{-11}$ m$^3$/C. The effect of different EC functionals on the flexoelectric coefficient can be found in Supplementary Table 1 and 2 and Supplementary Note 5.

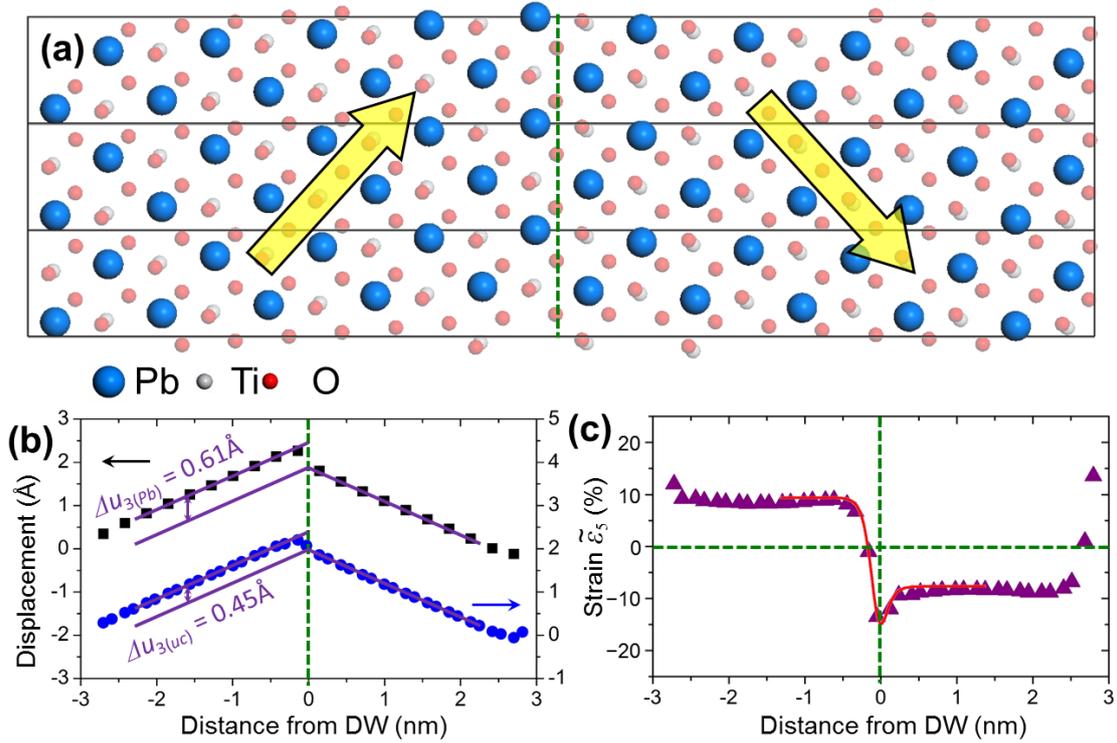

**Figure 3 | First-principles calculations to determine the flexoelectric coefficient $F_{11}$–$F_{12}$.** (**a**) The atomic model of 90 °DWs. (**b**) The distributions of Pb displacements and unit cell's displacements. (**c**) The distribution of the shear strain $\tilde{\varepsilon}_5$. The vertical green dashed lines mark the positions of DWs.

In our previous studies[12,20], offsets of the Pb sublattice were observed around 180 ° DWs in different PTO films, which were used to estimate the flexoelectric coefficient $F_{44}$. The same as the case of 90 ° DWs, the authentic offset across a 180 ° DW is determined by all ions. Thus, the technique of first-principles calculation is used to determine the positions of all ions and the authentic offset.

Similar to the study of the 90 °DW model, two EC functionals and two sets of lattice parameters were considered. The calculated Pb sublattice offset using the PBE-sol functional with ELP is 0.70 Å (Supplementary Fig. 7b), consistent with our previous experimental result (0.72 Å)[20]. The unit cell's offset is smaller than the Pb sublattice offset (Supplementary Fig. 7b), similar to the result of the 90 °DW model. The shear strain profile shows peaks and valleys at different DWs (Supplementary Fig. 7c), similar to the curve of the converse flexoelectricity in Fig. 1(e), since the shear strain induced by the electrostrictive effect can be neglected. We also compared the calculated and experimental Ti displacements relative to Pb centers away from the DW, and found their values are very close (both are about 0.17 Å), which demonstrates that our calculation results are reliable. Similar to the case of the 90 ° DW, the unit cell's offset ($\Delta u_3$) and the spontaneous polarization around a 180 ° DW are liked by the following equation (Supplementary Note 1):

$$\Delta u_3 = -2 F_{44} P_s \qquad (3)$$

The flexoelectric coefficient $F_{44}$ can thus be calculated as $-2.7 \times 10^{-11}$ m$^3$/C.

Currently, there are two classes of experimental methods quantifying flexoelectric coefficients. The first is applying inhomogeneous stress or electric field to macroscopic samples and recording the induced polarization or strain[21-26]. The second is evaluating flexocoupling strengths by the examination of microscopic local distortions via high-resolution TEM[12,27,28]. The flexoelectric coefficients obtained from macroscopic methods may contain surface contributions[29], while microscopic methods mainly concerned the static bulk contribution, which is the intrinsic one. The accuracy of previous microscopic methods, however, is limited by the resolution of TEM and complicated by the boundary conditions of complex domain patterns[27,28].

The advantage of our microscopic method lies on: Firstly, the adoption of first-principles calculations can make up the deficiency of TEM resolution; Secondly, the boundary condition of the simple DW model considered is very clear from the theoretical point of view. As a result, the flexoelectric coefficients obtained by our method should be more reliable. According to the LGD theory, the magnitudes of flexoelectric coefficients are proportional to the offset of unit cells, which can be determined accurately by doing a series of first-principles calculations (Supplementary Note 3). The sign of a flexoelectric coefficient is related with those of the polarization gradient and the induced shear strain around a DW, which are unambiguous from both first-principles calculations and the TEM atomic mapping. To sum up, the flexoelectric coefficients obtained from our method are very accurate in both magnitude and sign.

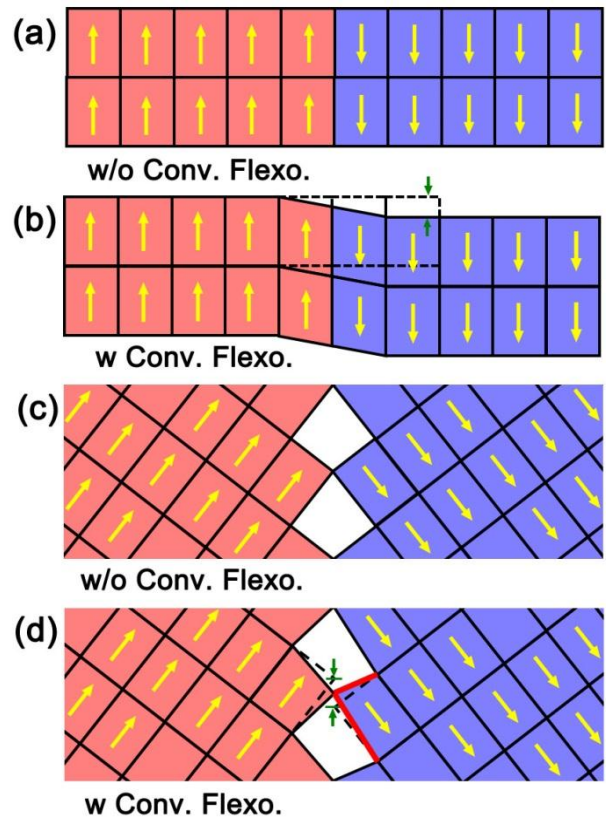

**Figure 4 | The schematic diagrams showing the effect of converse flexoelectricity on 180 ° and 90 ° DWs in tetragonal ferroelectrics.** The lattice offset is marked by green arrows in **b** and **d**. The asymmetric structure of 90 ° DWs, i.e. the abnormal maximal and minimal lattice parameters are marked by bold red lines in **d**.

Our results indicate that the converse flexoelectricity has direct relationship with the domain wall structures, which was largely neglected previously. If the converse flexoelectricity is not considered, there should be no lattice offset around 180 ° DWs and 90 ° DWs should be symmetric. It is the converse flexoelectricity that makes 180 ° DWs and 90 ° DWs in tetragonal ferroelectrics in their present shapes (Fig. 4). A natural extension is to explore the impact of converse flexoelectricity on other types of ferroelectric DWs, or other boundaries in ferroelectrics, such as morphotropic phase boundaries, antiphase boundaries, grain boundaries, and so on. The study about the converse flexoelectricity may not only help to understand many physical phenomena related with the gradients of electric field or polarization, but also pave new ways to the design of novel electromechanical devices.

**Methods**

**Thin film growth.** PTO thin films were grown on single crystal (001)-oriented $DyScO_3$ and (101)-oriented (La, Sr)(Al, Ta)$O_3$ substrates to obtain high-quality samples containing 90 °DWs. The growth technique was chosen as the pulsed laser deposition (PLD) with a KrF excimer laser ($\lambda$=248 nm) and the sintered PTO ceramic target (3 mol% Pb-riched) was used. Before the deposition, all substrates were pre-heated at 800 ℃ for 5 min to clean the substrates surface, and then cooled down to 700 ℃ to grow PTO thin films. The PTO target was pre-sputtered for 5 min. During the growth of PTO films, an oxygen pressure of 10 Pa, laser energy density of 2 J cm$^{-2}$ and a repetition rate of 4 Hz were used. After the deposition, the samples were kept at their growth temperature for 5 min and then cooled down to room temperature with a cooling rate of 5 ℃ min$^{-1}$ in an oxygen pressure of $3\times10^4$ Pa.

**TEM experiments.** The cross-sectional samples were prepared by slicing, gluing, grinding, dimpling and finally ion milling by Gatan PIPS. The final voltage of ion milling was less than 0.5 kV to reduce ion beam damage. HAADF-STEM images were recorded using Titan G$^2$ 60-300 microscope with a high-brightness field-emission gun, double aberration (Cs) correctors from CEOS and a monochromator operating at 300 kV. The beam convergence angle is 25 mrad and its resolution of STEM mode is up to 70 pm.

**First-principles calculations.** The atomic relaxations were performed by Vienna ab-initio simulation package (VASP)[30]. The LDA and PBE-sol EC functionals were used with the method of projector augmented-wave. The energy cutoff was chosen as 550 eV. O's 2s2p, Ti's 3s3p3d4s, and Pb's 5d6s6p electrons are treated as the valence electrons. The ionic relaxation was considered as convergent when the force on each ion is less than 2 meV/Å.

**Acknowledgements**

This work is supported by the Key Research Program of Frontier Sciences CAS (QYZDJ-SSW-JSC010) and the National Natural Science Foundation of China (No. 51571197 and 51671194). Y. L. T. acknowledges the support by the Youth Innovation Promotion Association CAS (No. 2016177). We are grateful to Mr. B. Wu and Mr. L. X. Yang of this lab for their technical support on the Titan platform of G2 60-300kV aberration-corrected scanning transmission electron microscope. Y. J. W. is grateful to Prof. A. N. Morozovska for the useful discussion of the LGD theory.

**Author contributions**

X.L.M. and Y.L.Z. conceived the project of interfacial characterization in oxides by using aberration-corrected STEM. Y.L.T. and Y.P.F. obtained high-resolution images and extracted atomic mappings. Y.J.W. analyzed atomic mappings via the converse flexoelectricity and did first-principles calculations to obtain flexoelectric coefficients. All authors contributed to the discussions and manuscript preparation.